# On a stochastic version of Lanchester's model of combat

Michael J Kearney [1] and Richard J Martin [2]

[1] Senate House, University of Surrey, Guildford, Surrey GU2 7XH, UK

[2] Department of Mathematics, Imperial College London,

South Kensington, London, SW7 2AZ, UK

*Abstract*

Lanchester's model of combat has certain deficiencies in its standard form arising from the neglect of the influence of random fluctuations. Several approaches to rectify this have been proposed and various results are scattered throughout the literature. Here, a discrete-time stochastic version, which is amenable to exact solution, is revisited with the aim of deriving key results within one setting. The exposition simplifies and provides refinements to earlier derivations and analysis.

## 1. Introduction

In its standard form, Lanchester's model of combat [1, 2] describes two groups engaged in an attritional battle whose populations $x(t)$ and $y(t)$ evolve according to,

$$\frac{dx}{dt} = -\frac{g(y)}{d(x,y)}; \qquad \frac{dy}{dt} = -\frac{f(x)}{d(x,y)}. \qquad (1)$$

Here, $f(x)$ and $g(y)$ are positive, strictly increasing functions representing the combat strengths of the two groups. A natural choice is $f(x) = ax$; $g(y) = by$, known as the 'square-law' case (see below as to why), but other choices are possible [3]. The 'screening function' $d(x, y)$ is often set equal to one. With complete generality, for any $d(x, y) > 0$, the solution trajectory $(x(t), y(t))$ satisfies,

$$F(x(t)) - G(y(t)) = C;$$
$$F(x) \equiv \int_0^x f(x') dx', \qquad G(y) \equiv \int_0^y g(y') dy' \qquad (2)$$

where $C \equiv F(M) - G(N)$ is a constant of the motion which depends on the initial values $x(0) = M$ and $y(0) = N$. The group whose population reduces to zero first is deemed to have lost the battle. This modelling framework can be employed to study competition in numerous areas beyond polemology (Lanchester's original area of interest), ranging from microeconomics through to ecology.

The structure of (2) already provides some interesting insights [3]. First, the screening function $d(x, y)$ affects the time to realize victory but *not* the trajectory (one can always transform the time variable to absorb it). Second, if $C > 0$ then the group whose population is labelled $x(t)$ is victorious (i.e. $y(t)$ will reach zero first) and there is a specified number of 'survivors' $S$ given by the implicit equation $F(S) = C$. Conversely, if $C < 0$ then the group whose population is labelled $y(t)$ is victorious and the number of survivors is given by $G(S) = -C$. Logically, therefore, a perfectly matched battle is one where $C = 0$. However, the model is clearly deficient as $C \to 0$ since the number of survivors vanishes. More realistically, if $C$ is 'small' stochastic effects cannot be ignored, since fluctuations eventually lead to one group or the other acquiring a decisive advantage which ultimately leads to victory.

An adaptation which allows for an element of randomness was studied in some detail in [4]. This built upon a simpler, discrete-time model presented in [5]. One can generalise the latter to suit our present purpose (see also [6, 7]) by defining a stochastic process on an integer state-space with transition probabilities,

$$P[(x, y) \to (x-1, y)] = \frac{g(y)}{f(x) + g(y)}$$

$$P[(x, y) \to (x, y-1)] = \frac{f(x)}{f(x) + g(y)}$$

(3)

starting at $(x, y) = (M, N)$ and terminating when either $x$ or $y$ reaches zero. The connection with the deterministic model (1) becomes clearer after making the formal choice $d(x, y) = f(x) + g(y)$. The fundamental quantity of interest in this model is the probability $P(M, f; N, g | S)$ that Group #1, whose initial population was $M$, is victorious with precisely $S$ survivors. From this, the overall probability of that group being victorious is given by $P(M, f; N, g) \equiv \Sigma_{S=1}^{M} P(M, f; N, g | S)$. Regarding notation, the probability that Group #2, whose initial population was $N$, is victorious with $S$ survivors is denoted $P(N, g; M, f | S)$, and clearly it must be the case that $P(M, f; N, g) = 1 - P(N, g; M, f)$. The analysis in [4] stops short of providing a general expression for $P(M, f; N, g | S)$, but it was derived in the context of diminishing urn models in [6, 7], building on the work in [8]. A *fully simplified* expression for $P(M, f; N, g)$ has not been provided to our knowledge.

Long ago, exact solutions for $P(M, f; N, g | S)$ and $P(M, f; N, g)$ for certain special cases of the process (3) were found and given in [9, 10]. One of these is the square-law case $f(x) = ax$; $g(y) = by$ which describes global, all-onto-all combat. The name is informed by how numerical superiority determines which group is victorious in the deterministic model, for which the relevant criterion is whether $C = \frac{1}{2} aM^2 - \frac{1}{2} bN^2$ is greater or less than zero. As a matter of historical interest, when $a = b = 1$ the square-law case reduces to the so-called 'OK Corral' model, which was defined quite independently in [11], analysed further in [5, 12] and comprehensively explored in [13, 14]. Inspired by the infamous gunfight in Tombstone, Arizona in 1881, the OK Corral model evocatively captures the essence of two groups engaged in random all-onto-all combat. We take the opportunity here to say that recognition for providing the solution to this model should also go to the authors of [9, 10]. Outside of their analyses, however, which rely heavily on the precise form of $f(x)$ and $g(y)$, less is known.

In this paper we show how to derive exact expressions for $P(M,f;N,g|S)$ and $P(M,f;N,g)$ for general $f(x)$ and $g(y)$. The primary motivation behind the work is to simplify and unify what is already known into a single framework. In that spirit, where results are simply extensions to existing results they are clearly presented as such. Our approach is to apply the basic embedding technique alluded to in [4, 5] but in a more direct manner, making no reference to the Friedman urn process and therefore avoiding the need for any time-reversal transformation. This more direct approach was articulated in [7] but the execution here is, in several respects, rather different and leads to alternative representations of certain results which prove to be of great utility. The embedding method enables one to express $P(M,f;N,g|S)$ and $P(M,f;N,g)$ in terms of integrals involving the transition probabilities of a pure death process. These integrals reduce quickly to give compact results which agree with those derived in [9, 10] for the special cases using different techniques. The integral representations also naturally lead to approximations which are, in practical terms, often accurate even for quite small values of $M,N$.

**2. Exact solutions**

The time evolution of the process may be thought of as a decreasing directed path on a square lattice, whose axes label the populations of the two groups and where individual steps are taken with probabilities given by (3); see figure 1. It is evident that $P(M,f;N,g|S)$ can be expressed as the sum of the probabilistic weights of all possible paths which start at the point $(M,N)$ and end at the point $(S,0)$. To carry out this procedure we employ a technique, described in more detail in [5, 7, 15], the origins of which may be traced back to the work in [16] on embedding urn schemes into continuous time Markov branching processes. In what follows we use the shorthand notation $f(k) \equiv f_k$ and $g(k) \equiv g_k$ for $k > 0$.

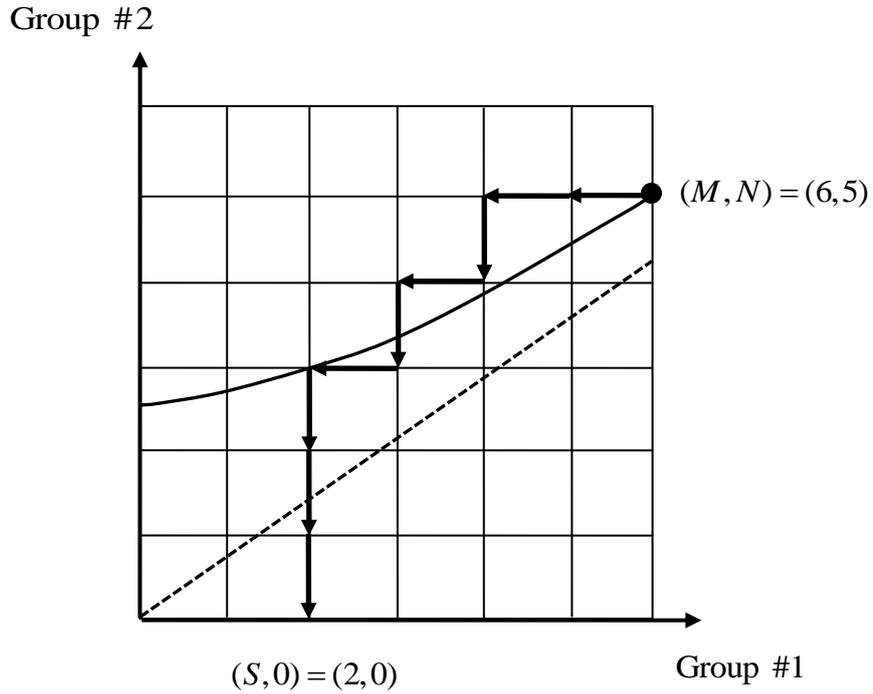

**Figure 1**: A step by step trajectory of the process for a square-law battle with $a=1, b=2$, starting at $(M,N)=(6,5)$ and ending at $(S,0)=(2,0)$. The straight dashed line represents the boundary marking equal combat strength ($C=0$). In this example, the weaker group at the outset (Group #1, for which the win probability is $\approx 0.31$) is victorious. The curved line represents the deterministic solution trajectory, for which the stronger group at the outset is always victorious.

Let $X(t)$ be a continuous-time death process on the non-negative integers with state-dependent transition rates $f_k^{-1}$ and an absorbing state at $k = 0$, and suppose $X(0) = M$. We denote the probability that $X(t)$ is in state $k \leq M$ at time $t$ by $P_f(M,k;t)$ which satisfies the master equation,

$$\frac{d}{dt} P_f(M,k;t) = f_{k+1}^{-1} P_f(M,k+1;t) - f_k^{-1} P_f(M,k;t)$$

with $f_0^{-1} \equiv 0$. This is straightforward to solve by Laplace transforms and residue calculus (recall that $f(x)$ is strictly increasing so $f_i \neq f_j$ unless $i = j$);

$$P_f(M,k;t) = f_k \sum_{j=k}^{M} \frac{1}{f_j} \left\{ \prod_{\substack{i=k \\ i \neq j}}^{M} \frac{f_j}{f_j - f_i} \right\} e^{-t/f_j}. \tag{4}$$

This simplifies when $k = M$ to give $P_f(M,M;t) = e^{-t/f_M}$. The expression holds for $k = 0$ in the sense that one may set $k = 0$ and take the limit $f_0 \to \infty$. This yields,

$$P_f(M,0;t) = 1 - \sum_{j=1}^{M} \left\{ \prod_{\substack{i=1 \\ i \neq j}}^{M} \frac{f_j}{f_j - f_i} \right\} e^{-t/f_j}. \tag{5}$$

The probability density function $p_f(M,k;T)$ for the first passage time $T$ for $X(t)$ to reach state $k < M$ is simply the transition rate from state $k+1$ multiplied by the probability of being in state $k+1$ at time $T$, thus,

$$p_f(M,k;T) = f_{k+1}^{-1} P_f(M,k+1;T)$$

(6)

$$= \sum_{j=k+1}^{M} \frac{1}{f_j} \left\{ \prod_{\substack{i=k+1 \\ i \neq j}}^{M} \frac{f_j}{f_j - f_i} \right\} e^{-T/f_j}.$$

When $k = M-1$ this simplifies to give $p_f(M, M-1;T) = f_M^{-1} e^{-T/f_M}$.

Together with $X(t)$, let us consider an independent death process $Y(t)$ with $Y(0) = N$ and state-dependent transition rates $g_k^{-1}$. The probability that $Y(t)$ makes the jump from $N$ to $N-1$ before $X(t)$ makes the jump from $M$ to $M-1$ is straightforward to evaluate;

$$P[(M,N) \to (M, N-1)] = \int_0^\infty p_g(N, N-1;T) P_f(M,M;T) \, dT$$

$$= \frac{f_M}{f_M + g_N}$$

and, conversely, $P[(M,N) \to (M-1,N)] = g_N / (g_N + f_M)$. With reference to (3), this generates the correct transition probabilities for the first step of the process. The basic idea then comes from recognising that, since the processes $X(t)$ and $Y(t)$ are memoryless, by integrating over all possible realisations one can effectively sum over sets of lattice paths, the probabilistic weights of which are automatically generated and correctly counted [15]. The independence of $X(t)$ and $Y(t)$ means the model is effectively decoupled; see also [5, 7].

One can therefore evaluate $P(M, f; N, g | S)$ by integrating over all realisations in which, at the time the process $Y(t)$ makes the transition to state $0$, the process $X(t)$ is in a given state $S > 0$;

$$P(M, f; N, g | S) = \int_0^\infty p_g(N, 0; T) P_f(M, S; T) \, dT. \tag{7}$$

Using (4) and (6) one obtains, after carrying out the integral,

$$P(M, f; N, g | S) = f_S \sum_{\ell=1}^{N} \left\{ \prod_{\substack{u=1 \\ u \neq \ell}}^{N} \frac{g_\ell}{g_\ell - g_u} \right\} \sum_{j=S}^{M} \frac{1}{f_j} \left\{ \prod_{\substack{i=S \\ i \neq j}}^{M} \frac{f_j}{f_j - f_i} \right\} \frac{f_j}{f_j + g_\ell}.$$

This looks cumbersome, but the summation over $\ell$ can be performed exactly using the following partial fraction identity (noting $g_i \neq g_j$ unless $i = j$);

$$\prod_{\ell=k}^{N} \frac{1}{1 + g_\ell z} \equiv \sum_{\ell=k}^{N} \left\{ \prod_{\substack{u=k \\ u \neq \ell}}^{N} \frac{g_\ell}{g_\ell - g_u} \right\} \frac{1}{1 + g_\ell z} \tag{8}$$

with the assignment $z = f_j^{-1}$, leading to the final result, valid for $S > 0$,

$$P(M, f; N, g | S) = f_S \sum_{j=S}^{M} \frac{1}{f_j} \left\{ \prod_{\substack{i=S \\ i \neq j}}^{M} \frac{f_j}{f_j - f_i} \right\} \left\{ \prod_{\ell=1}^{N} \frac{f_j}{f_j + g_\ell} \right\}. \tag{9}$$

This expression is *exact*, and, with modest effort, one can see that it is the same as the expressions given in [6, 7]. From (9) one can evaluate $P(M,f;N,g)$ by summing over $S$. This can be done by reversing the order of the summations and using the following identity which can be proved by induction on $k$;

$$\sum_{S=k}^{j} \frac{f_S}{f_j} \left\{ \prod_{i=S}^{j-1} \frac{f_j}{f_j - f_i} \right\} \equiv \prod_{i=k}^{j-1} \frac{f_j}{f_j - f_i}.$$

The result is,

$$P(M,f;N,g) = \sum_{j=1}^{M} \left\{ \prod_{\substack{i=1 \\ i \neq j}}^{M} \frac{f_j}{f_j - f_i} \right\} \left\{ \prod_{\ell=1}^{N} \frac{f_j}{f_j + g_\ell} \right\}. \qquad (10)$$

This expression, as far as we are aware, has not been presented before. Alternatively, one can derive (10) by integrating over all realisations in which, at the time the process $Y(t)$ makes the transition to state $0$, the process $X(t)$ is *not* in state $0$;

$$P(M,f;N,g) = \int_0^\infty p_g(N,0;T)\left[1 - P_f(M,0;T)\right] dT. \qquad (11)$$

Using (5) and (6) one obtains after carrying out the integral,

$$P(M,f;N,g) = \sum_{\ell=1}^{N} \left\{ \prod_{\substack{u=1 \\ u \neq \ell}}^{N} \frac{g_\ell}{g_\ell - g_u} \right\} \sum_{j=1}^{M} \left\{ \prod_{\substack{i=1 \\ i \neq j}}^{M} \frac{f_j}{f_j - f_i} \right\} \frac{f_j}{f_j + g_\ell}.$$

Again, the summation over $\ell$ can be performed using (8) leading to (10). Together with (9), one has the two key results from which other statistical quantities related to the process can be calculated.

With reference to (3), and exploiting the directional nature of the process, it is clear that $P(M, f; N, g)$ must obey the recurrence equation,

$$P(M, f; N, g) = \frac{g_N}{g_N + f_M} P(M-1, f; N, g)$$

$$+ \frac{f_M}{f_M + g_N} P(M, f; N-1, g) \quad (12)$$

with boundary conditions $P(M, f; 0, g) = 1$ and $P(0, f; N, g) = 0$. One may show that (10) satisfies (12) as required. The symmetry $P(M, f; N, g) = 1 - P(N, g; M, f)$ is hard to see from (10) but may be established from (11) since,

$$P(M, f; N, g) = \int_0^\infty p_g(N, 0; T) dT - \int_0^\infty p_g(N, 0; T) P_f(M, 0; T) \, dT$$

$$= 1 - P(N, g; M, f).$$

The final step recognises that the second integral is over realisations for which, at the time the process $Y(t)$ makes the transition to state $0$, the process $X(t)$ is *already* in state $0$. All such realisations have the property that at the time the process $X(t)$ reached state $0$, $Y(t)$ was greater than $0$ almost surely; hence the group whose initial population was $N$ wins. This train of logic leads to an alternative expression,

$$P(M,f;N,g) = \int_0^\infty p_f(M,0;T) P_g(N,0;T)\, dT \qquad (13)$$

$$= \int_0^\infty p_f(M,0;T) \int_0^T p_g(N,0;T')\, dT'dT$$

where we have used the fact that since the state $k = 0$ is absorbing,

$$P_g(N,0;T) = \int_0^T p_g(N,0,T')\, dT'.$$

The symmetry $P(M,f;N,g) = 1 - P(N,g;M,f)$ follows from (13) by integrating by parts. The criterion for a perfectly matched battle is simply $P(M,f;N,g) = \tfrac{1}{2}$.

For the square-law case one has $f(x) = ax$; $g(y) = by$, whereupon the results (9) and (10) can be simplified to,

$$P(M,f;N,g|S) = S\left(\frac{a}{b}\right)^N \sum_{j=S}^M \frac{(-1)^{M-j} j^{M+N-S-1}}{(j-S)!(M-j)!} \frac{\Gamma(aj/b+1)}{\Gamma(N+aj/b+1)}$$

$$P(M,f;N,g) = \left(\frac{a}{b}\right)^N \sum_{j=1}^M \frac{(-1)^{M-j} j^{M+N}}{(M-j)!j!} \frac{\Gamma(aj/b+1)}{\Gamma(N+aj/b+1)}$$

where $\Gamma(z)$ is the Gamma function. These are precisely the results given in [9, 10], where they are derived after a challenging analysis from (12) and related expressions using a technique which has restricted applicability. As mentioned in the Introduction, by setting $a = b = 1$ the OK Corral model is recovered. This has been analysed in isolation and the relevant cases have been obtained independently [13, 14].

## 3. Large scale behaviour

Although (9) and (10) are exact, they are computationally intensive to evaluate at large scale. Fortunately, the integral representations (7) and (13) provide an ideal starting point for making accurate approximations. This is based on the fact that, provided the growth of the function $f(x)$ (and in turn $g(y)$) is no faster than polynomial (a weak restriction in practice), the probability density function (6) is asymptotically normal as $M \to \infty$ (see e.g. [15]), and in particular for $k = 0$,

$$p_f(M,0;T) \approx \frac{1}{\sqrt{2\pi \mathcal{F}(M)}} \exp\left\{-\frac{(T-F(M))^2}{2\mathcal{F}(M)}\right\} \qquad (14)$$

where $F(M)$ is given by (2) and,

$$\mathcal{F}(M) \equiv \int_0^M f(x')^2 \, dx'.$$

This is simply a consequence of the central limit theorem for independent but non-identically distributed random variables, noting that the mean and variance of the first passage time $T$ are given by,

$$\mathrm{E}[T] = \sum_{j=1}^M f_j \approx F(M); \qquad \mathrm{Var}[T] = \sum_{j=1}^M f_j^2 \approx \mathcal{F}(M).$$

Using (13) in conjunction with (14), as well as its natural extension to approximate $p_g(N,0;T)$ in terms of $G(N)$ and $\mathcal{G}(N) \equiv \int_0^N g(y')^2\, dy'$, and carrying out one of the integrals explicitly, one can obtain an estimate for $P(M,f;N,g)$;

$$P(M,f;N,g) \approx \Phi(\theta) \equiv \frac{1}{\sqrt{2\pi}} \int_{-\infty}^{\theta} e^{-t^2/2}\, dt$$

$$\theta(M,f;N,g) = \frac{F(M) - G(N)}{\sqrt{\mathcal{F}(M) + \mathcal{G}(N)}}.$$

(15)

This result was given in [4], but the derivation here is somewhat distinct. It should be noted from (15) that $\theta(M,f;N,g) = -\theta(N,g;M,f)$ which ensures that $P(M,f;N,g) = 1 - P(N,g;M,f)$. The condition for a perfectly matched battle is simply $\theta = 0$ which implies $F(M) = G(N)$ or $C = 0$. This is precisely as deduced from (2) but now it has a much broader interpretation.

The accuracy of (15) gets better as $M, N \to \infty$, but it can provide good answers even for small values of $M, N$. For example, for a square-law battle with $a = 1, b = 1$, starting at $(M,N) = (12,10)$, the exact probability according to (10) of the larger group winning is $= 0.768...$, whereas the approximate probability according to (15) is $\approx 0.77$. Of course, one should not push things too far at small scale; e.g., with $a = 1, b = 2$ and starting at $(M,N) = (12,8)$, the exact probability of the larger group winning is $= 0.575...$, whereas the approximate probability is $\approx 0.60$.

Concerning the number of survivors, one can estimate $P(M,f;N,g|S)$ using (7) by writing $P_f(M,S;T) \approx f(S)p_f(M,S;T)$ and replacing $F(M)$ in (14) with $F(M)-F(S)$. After carrying out the integral in (7) one has,

$$P(M,f;N,g|S) \approx \frac{f(S)}{\sqrt{2\pi(\mathcal{F}(M)+\mathcal{G}(N))}} \times \exp\left(-\frac{(G(N)-F(M)+F(S))^2}{2(\mathcal{F}(M)+\mathcal{G}(N))}\right). \quad (16)$$

Noting that $f(S) \equiv dF(S)/dS$ one has the consistency check,

$$P(M,f;N,g) = \sum_{S=1}^{M} P(M,f;N,g|S)$$

$$\approx \int_0^\infty P(M,f;N,g|S)\,dS = \Phi(\theta).$$

As far as we are aware (16) has not been presented before as such. From it one can calculate many quantities relating to the size of the surviving population. For example, the expected number of survivors $E_{\#1}[S]$, *given* that Group #1 whose initial population is $M$ wins, is as follows,

$$E_{\#1}[S] \equiv \sum_{S=1}^{M} S \frac{P(M,f;N,g|S)}{P(M,f;N,g)} \approx \frac{1}{\Phi(\theta)\sqrt{2\pi}} \int_0^\infty q(z) e^{-(z-\theta)^2/2}\,dz$$

$$q(z) = F^{-1}\left[z\sqrt{\mathcal{F}(M)+\mathcal{G}(N)}\right]. \quad (17)$$

If the level of combat superiority is such that Group #1 is overwhelmingly likely to be victorious, which implies $\theta \gg 1$ or $\Phi(\theta) \approx 1$, then the integrand of (17) is dominated by contributions in the vicinity of $z = \theta$. It may thus be approximated as,

$$\mathrm{E}_{\#1}[S] \approx q(\theta) = F^{-1}(F(M) - G(N)) = F^{-1}(C).$$

This is the same as the number of survivors in the deterministic model. Moreover, one may show that the variance $\mathrm{Var}_{\#1}[S] \approx \left(\mathrm{d}q(z)/\mathrm{d}z\big|_{z=\theta}\right)^2 \ll q(\theta)^2$, which implies that the number of survivors in any given battle is likely to be close (in the sense of a small coefficient of variation) to the deterministic result. This is intuitively appealing, since stochastic effects will be minimal if the odds of winning are overwhelming.

On the other hand, for a closely matched battle with $|\theta| \ll 1$ one needs a case-specific calculation. By way of illustration, for the square-law case with $r = \sqrt{b/a}$ whose initial populations satisfy $|M - rN| \ll \sqrt{M + rN}$, which is sufficient to ensure that $|\theta| \ll 1$, one has from (17),

$$\mathrm{E}_{\#1}[S] \approx K(1+r)^{1/4}(M+rN)^{3/4}\left[1 + \frac{L}{\sqrt{1+r}}\left(\frac{M-rN}{\sqrt{M+rN}}\right) + ...\right]$$

$$K = \frac{\Gamma(\tfrac{3}{4})}{3^{1/4}\sqrt{\pi}} = 0.525...; \qquad L = 2\sqrt{3}\left(\frac{\Gamma(\tfrac{5}{4})}{\Gamma(\tfrac{3}{4})} - \frac{1}{\sqrt{\pi}}\right) = 0.607....$$

A key observation is that the expected number of survivors does not scale linearly with the size of the populations, but rather with exponent $\tfrac{3}{4}$. When $r = 1$ we recover the OK Corral model, wherein for a perfectly matched battle with $M = N$ the expected number of survivors is given to leading order by $2KM^{3/4}$ [11, 12].

It is not our intention here to exhaustively explore different modelling options. However, one can easily go beyond the square-law case by characterising the combat strengths of the two groups with a choice such as the following which features saturation at large scale;

$$f(x) = \frac{a}{\alpha}\tanh(\alpha x); \qquad g(y) = \frac{b}{\beta}\tanh(\beta y).$$

From this it follows in turn that,

$$F(M) = \frac{a}{\alpha^2}\log\cosh(\alpha M); \qquad G(N) = \frac{b}{\beta^2}\log\cosh(\beta N)$$

$$\mathcal{F}(M) = \frac{a^2 M}{\alpha^2} - \frac{a^2}{\alpha^3}\tanh(\alpha M); \qquad \mathcal{G}(N) = \frac{b^2 N}{\beta^2} - \frac{b^2}{\beta^3}\tanh(\beta N).$$

These reduce to the square-law case in situations where $M \ll \alpha^{-1}$ and $N \ll \beta^{-1}$, but saturate to the so-called 'linear-law' case when $M \gg \alpha^{-1}$ and $N \gg \beta^{-1}$. The latter is characteristic of sequential one-onto-one combat rather than all-onto-all combat; see e.g. the discussion in [3, 9, 10]. Through careful selection of parameters one can use these expressions in conjunction with earlier results to explore a wide range of intermediary (mixed) models.

We close with a tangential remark. For the weaker group to win there must be a cross-over point during the battle (see figure 1) where both groups are approximately equally matched, whereupon either group can ultimately win with probability $\approx \frac{1}{2}$. Thus one may write down, with increasing accuracy at large scale, that $\Phi(-|\theta|) \approx P_{Eq} \times \frac{1}{2}$ or $P_{Eq} \approx 2\Phi(-|\theta|)$, where $P_{Eq}$ is the probability that the two groups become equally matched at some juncture [15]. It then follows that the probability of the stronger group not just winning but maintaining a persistent advantage *at all stages* of the battle is given using (15) by,

$$P_> \equiv 1 - P_{Eq} \approx \sqrt{\frac{2}{\pi}} \int_0^{|\theta|} e^{-t^2/2} \, dt. \qquad (18)$$

One might wish for this to happen with a high degree of confidence; knowledge of $\theta$ through (15) together with (18) provides valuable insights.

## 4. Conclusions

We have presented a solution to a stochastic version of Lanchester's model of combat, the key results being (7), (9) and (10), (13), from which other statistically relevant quantities may be calculated. Our principal aim has been to pull together results derived by different methods and in the context of other models. The construct allows for key asymptotic results to be derived straightforwardly. These shed important light on the nature of the model and the impact of scale and initial imbalance in shaping the subsequent dynamics.